# Spin readout via spin-to-charge conversion in bulk diamond nitrogen-vacancy ensembles

Harishankar Jayakumar[1], Siddharth Dhomkar[1], Jacob Henshaw[1,2], and Carlos A. Meriles[1,2,†]

[1]*Dept. of Physics, CUNY-City College of New York, New York, NY 10031, USA.*
[2]*CUNY-Graduate Center, New York, NY 10016, USA.*

We demonstrate optical readout of ensembles of nitrogen-vacancy(NV) center spins in a bulk diamond sample via spin-to-charge conversion. A high power 594 nm laser is utilized to selectively ionize these paramagnetic defects in the $m_S = 0$ spin state with a contrast of up to 12%. In comparison with the conventional 520 nm spin readout, spin-to-charge-conversion-based readout provides higher signal-to-noise ratio, with tenfold sensing measurement speedup for millisecond long pulse sequences. This level of performance was achieved for an NV⁻ ionization of only 25%, limited by the ionization and readout laser powers. These observations pave the way to a range of high-sensitivity metrology applications where the use of NV⁻ ensembles in bulk diamond has proven useful, including sensing and imaging of target materials overlaid on the diamond surface.

The negatively charged nitrogen-vacancy (NV⁻) center in diamond has recently emerged as a versatile room-temperature nanoscale optical sensor of magnetic field[1–3], electric field[4], pressure[5] and temperature[6]. To improve detection sensitivity and spatial resolution, an extensive effort has been devoted to extending the electron spin coherence time[7,8], augmenting the photo-luminescence (PL) collection efficiency[9], and growing the number of NVs in the detection volume[9,10]. Sensitive spin spectroscopy has also been performed with T1 relaxometry measurements[11]. Further, magnetic field sensing has been exploited to detect nuclear spins in nearby molecules[12] and proteins[13], and to image vortices in superconductors[14] and domain walls in ferromagnetic materials[15]. To meet various sensing needs, experiments have been performed with single and ensemble NV centers in bulk diamond, near the surface of bulk crystals[13], or in nanodiamonds[1] and nano-photonic-structures[16].

Among the keys to the success of the NV⁻ as a spin-based sensor is the ability to optically readout its spin polarization, encoded as the difference in the photoluminescence intensity when the spin is in the $m_S = 0$ or $\pm 1$ states of the electronic ground state triplet[17] (see Fig. 1a). The intensity difference or 'readout contrast' is a direct result of the spin-dependent optical cycling frequency of electrons between the ground and excited triplet states. Optical excitation from the ground $m_S = 0$ state produces efficient photon emssion, with spin populations predominantly remaining within the triplet manifold. On the contrary, optical excitation from the ground $m_S = \pm 1$ spin states is prone to intersystem crossing (upon which the electron is shelved in a metastable singlet state), thereby reducing the photon emission rate; the ensuing spin contrast in the collected PL reaches about 30% in individual NVs and up to 20% in ensembles.

Importantly, the very mechanisms leading to NV⁻ spin initialization also limit the time window for photon-collection to only ~300 ns, on average leading to less-than-one photon per observation in a typical confocal setup. Shields et al.[18] and Hopper et al.[19] demonstrated that the readout times in single NVs can be extended by transforming the charge state from negative to neutral, conditional on the initial NV⁻ spin state. This strategy can substantially improves detection sensitivity, particularly in sensing experiments with long wait times between successive repetitions.

Understandably, there is interest in expanding this single-NV work to ensembles in type 1b crystals, attractive for high-sensitivity detection in applications where high spatial resolution (e.g., <100 nm) is not critical. Thus far, however, demonstrations have proven elusive, in part because multi-defect charge processes absent in ultra-pure type 2a diamond tend to complicate the ionization/recombination dynamics of the NV[20,21]. Recently, charge state readout and spin-to-charge conversion (SCC) were reported for NV ensembles in diamond nanocrystals[22], but NVs in high-pressure-high-temperature (HPHT) nm-size particles are prone to various non-radiative transition pathways, hence obscuring the dynamics at play. Further, the finite volume of diamond nanocrystals ensures the illumination intensity — and thus the charge conversion rates at play — are uniform, a condition impossible to meet when the ensemble extends beyond the beam size. In particular, the excitation and ionization volumes — respectively governed by single- and two-photon processes — are different in a bulk crystal, thus casting doubts on the effectiveness of spin-to-charge conversion protocols, inherently designed to articulate both processes. In spite of these complications, here we report spin readout via spin-to-charge conversion in high-density bulk NV ensembles; compared to conventional spin readout, we attain up to a tenfold enhancement, limited by our present experimental conditions.

For this study, we use a 0.2 mm thick, [100], type 1b diamond with a nominal nitrogen concentration of 1 ppm purchased from Delaware Diamond Knives. We investigate NV ensembles at a depth of about 20 μm from the sample surface via a home-built laser-scanning microscope with a 1.3 NA oil immersion objective; the beam diameter is ~300 nm corresponding to a detection volume $V_d$ of approximately 0.07 μm³. From the observed fluorescence, we conclude the number of NVs within $V_d$ amounts to ~30, hence yielding an NV concentration of ~3 ppb. To deliver microwave (MW), we use a 20 μm copper wire overlaid on the crystal surface. NV initialization and conventional spin readout is performed with a

[†]E-mail: cmeriles@ccny.cuny.edu.

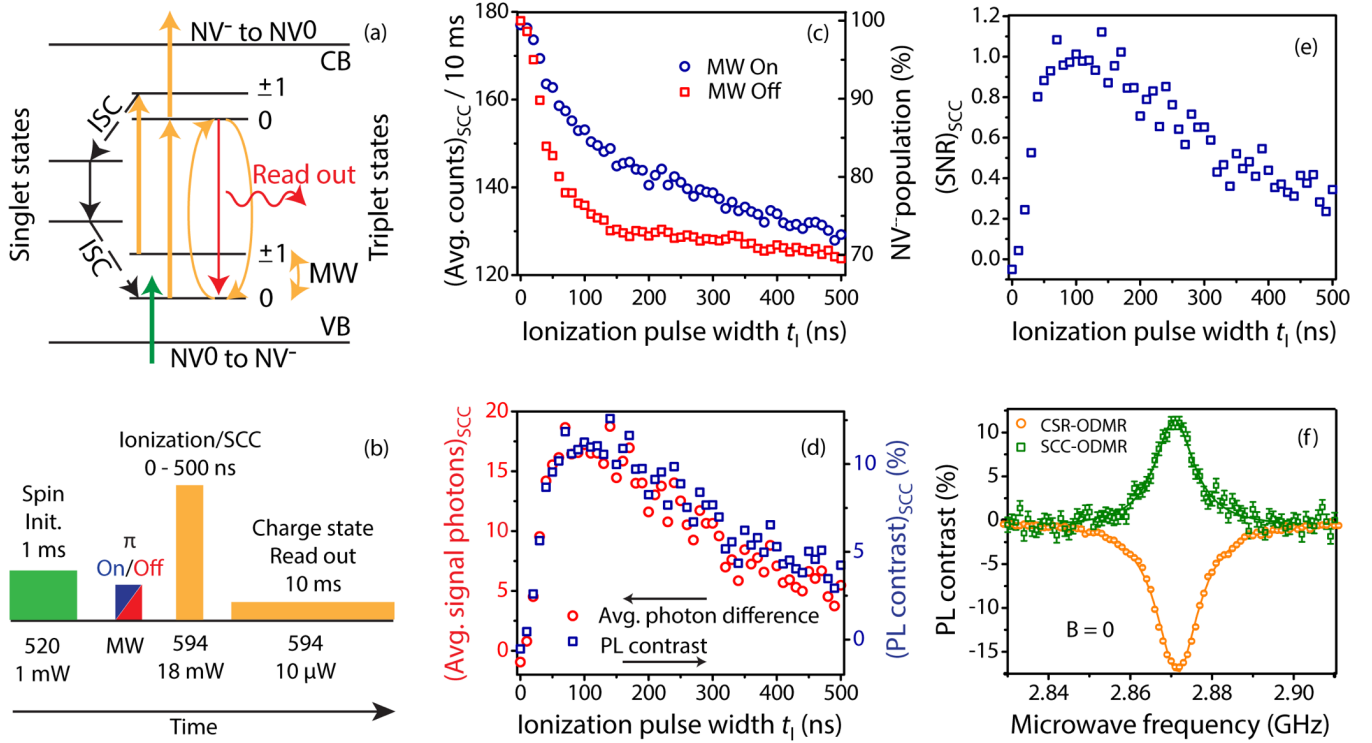

**Figure 1**: Fundamentals of SCC. (a) NV⁻ energy level diagram and schematics of spin-to-charge conversion. Charge and spin initialization is carried out via 520 nm laser excitation (green arrow). We ionize and readout the NV charge state via strong and weak 594 nm laser light (thick and thin orange arrows, respectively). The wavy, red arrow denotes emitted photons during detection. MW: Microwave. ISC: Inter-system crossing. CB: Conduction band. VB: Valence band. (b) Pulse sequence for SCC-based spin-state measurement. (c) Average number of photons acquired by the 10 ms readout pulse after the spin-to-charge conversion, with and without the MW π-pulse. (d) Average number of signal photons and contrast for a 10 ms readout pulse for varying ionization pulse widths. (e) Signal-to-noise ratio of the SCC measurement for varying ionization pulse widths. (f) Comparison of the pulsed ODMR signals acquired with spin-to-charge conversion (SCC) and conventional spin readouts (CSR) at zero external magnetic field ($B = 0G$). Solid lines are Lorenzian peak function fits to the data and the error bars are for the photon shot noise.

520 nm pulsed diode laser. NV⁻ ionization, on the other hand, relies on high-power 594 nm laser illumination, pulsed with an acousto-optic modulator (AOM) in the single pass configuration; a weaker, 10 μW laser serves as the readout beam. A single photon detector module in conjunction with a single mode optical collection fiber is used for high resolution readout. We collect the NV⁻ phonon side band emission upon crossing a long-pass 650 nm filter. All required pulses are generated through a commercial field-programmable gate array (FPGA).

The spin-to-charge conversion protocol we implement herein is schematically presented with the aid of the energy level diagram in Fig. 1a. The charge and the spin state of the NVs are initialized by a 1 mW 520 nm laser, with the spin state manipulated via MW pulses. The ionization pulse converts the charge state from NV⁻ to NV⁰ via a two-step, two-photon process active predominantly when the initial spin state is $m_S = 0$. In the first step of the ionization process, the electron is taken to the excited state where it has a comparatively longer radiative lifetime of 13 ns, and thus a greater, laser-power-dependent probability of being ionized by a second photon. On the other hand, if the initial spin projection is $m_S = \pm 1$, the excited state rapidly undergoes intersystem crossing and is shelved for 300 ns in the ground singlet, a state with a low two-photon ionization cross section. Thus, the efficiency of the spin-to-charge conversion depends on (i) the spin selectivity of the intersystem crossing, (ii) the ionization probabilities of the triplet and singlet states, (iii) the ionization laser power and pulse width, and (iv) the recombination probability. While the intrinsic properties of the NV obviously play a key role, the ultimate efficiency can be optimized through a careful choice of laser wavelength, power, and pulse width. In our experiments, we implement SCC via an 18 mW, 594 nm laser of varying pulse width and with ionization-to-recombination ratio[23,24] of 7:1. Following the spin-to-charge conversion, the resulting charge state is readout with the 10 μW 594 nm laser beam. The readout time can be chosen considering the trade-off between laser power, sequence length, and ionization rate.

The pulse sequence used for the experimental realization of SCC (Fig. 1b) consists of a 520 nm initialization laser pulse (1 mW, 1 ms), a MW π-pulse for spin inversion, an 18 mW, 594 nm, laser ionization pulse of varying duration, and a weaker 594 nm readout pulse (10 μW, 10 ms). The ionization pulse width was swept in the range 0-500 ns to determine the highest possible signal-to-noise ratio (SNR). A lower photon count (and hence an increased ionization rate) is observed for $m_S = 0$ (when the MW is off) due to spin-to-charge conversion (Fig. 1c). About 25% (14%) of NVs are converted to NV⁰ within the first 100 ns of the ionization pulse for $m_S = 0$ ($m_S = \pm 1$). From here, we define the average signal photon (SP) count as

$$SP = C_{\text{MW On}} - C_{\text{MW Off}},$$

where $C_{\text{MW On, Off}}$ are the average counts shown in Fig. 1c. The SCC photo-luminescence (PL) contrast is given by

$$PL\ Contrast = 1 - \frac{C_{\text{MW Off}}}{C_{\text{MW On}}}.$$

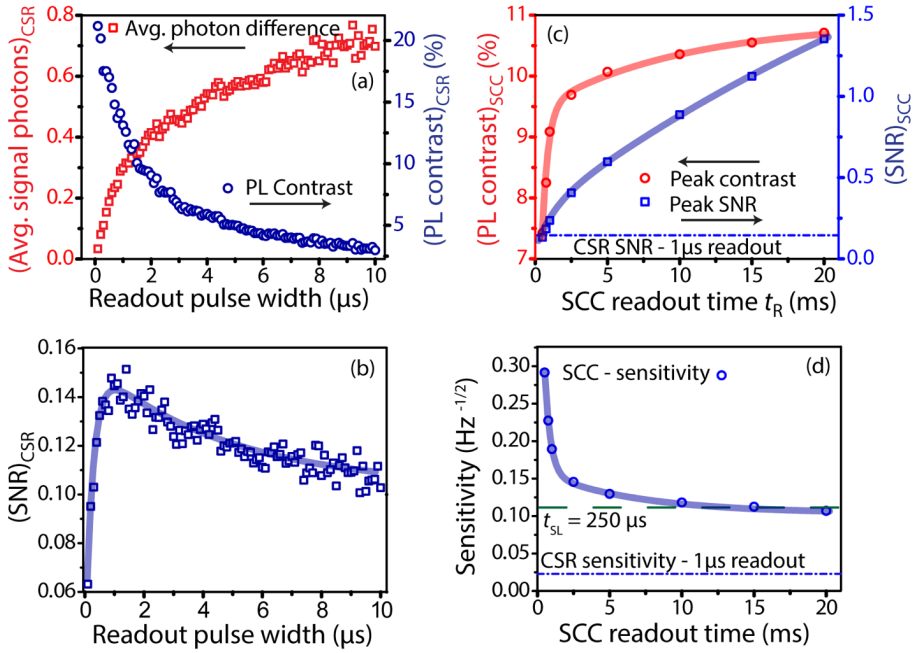

**Figure 2**: Comparison between SCC and standard spin readout. (a) Average number of signal photons and contrast, and (b) SNR for conventional 520 nm spin state read out without spin-to-charge conversion, as a function of readout pulse width. (c) PL contrast (SNR) for SCC readout of variable duration (red and blue traces, left and right vertical scales, respectively) ; for comparison, the plot includes the SNR from 520 nm, 1 μs readout (dash-dot blue line). (d) A similar comparison for the readout sensitivity. The dash-dot and dashed lines are sensitivities for 520 nm, 1us readout assuming sequence lengths of $t_{SL} = 0$ and 250 μs, respectively. Solid lines in (b), (c), and (d) are guides to the eye.

The SNR is calculated with the photon shot noise as,

$$SNR = \frac{SP}{\sqrt{(C_{\text{MW On}} + C_{\text{MW Off}})}}.$$

The difference in the $m_S = 0$ and $m_S = \pm 1$ count rates peaks at 17 photons corresponding to a maximum SCC PL contrast of 11% (Fig. 1d), compared to 13% obtained with the conventional spin state readout in this sample. We attain a maximum SNR of 1.0 for a ioniozation pulse duration of about 100 ns (Fig. 1e). The optically detected magnetic resonance (ODMR) signal acquired with a 100 ns ionization pulse is shown in Fig. 1f; note that in contrast to the conventional ODMR measurement, the normalized SCC PL intensity must now increase at the resonance frequencies, as observed experimentally.

We next characterize the SCC performance as a spin readout protocol relative to the conventional spin readout (CSR) at 520 nm. To use the latter as a reference, we first determine the optimum conditions by sweeping the CSR pulse width from 0 to 10 μs, following a 1 mW, 10 μs spin polarization pulse. While the number of collected photons increases with the readout time (Fig. 2a), the CSR PL contrast decreases due to spin re-initialization into $m_S = 0$. A peak SNR of 0.15 is achieved for 1 μs readout, with an average of 0.3 signal photons and 13% contrast (Fig. 2b).

Results from observations using SCC are presented in Fig. 2c for a variable readout time. Unlike conventional readout, we find that the SCC PL contrast and SNR increase with longer readout times. The characteristic time constants — within the ms range in both cases — are set by the readout beam power, limited in these experiments to only 10 μW. Fig. 2d displays the sensitivity $\eta$ of the spin state measurement defined as

$$\eta = \frac{\sqrt{t_T}}{\text{SNR}},$$

where the total protocol time $t_T$ is here given by

$$t_T = t_I + t_{SL} + t_R,$$

and $t_I$ denotes the initialization time, $t_{SL}$ is the sequence length, and $t_R$ indicates the readout time. Using $t_R = 20$ ms, the SCC protocol reaches a limit sensitivity of about 0.1 Hz$^{-1/2}$, considerably poorer than the sensitivity of conventional 520 nm readout (0.025 Hz$^{-1/2}$, as highlighted by the blue dash-dot line in Fig. 2d). Aside from the fact that the SCC ionization pulse power available to us is not sufficient to get the maximum ionization/contrast (see below), a major limitation arises from the 1 ms pulse required to properly initialize the NV charge state, much longer than the 10 μs pulse typical in conventional spin readout. In practical sensing applications, however, successive spin readouts are separated in time by the intrinsic duration of the measurement protocol (or 'sequence length' $t_{SL}$), typically in the range of hundreds of microseconds to a few milliseconds. Even under these suboptimal conditions, the present SCC protocol starts to be advantageous relative to CSR for sequences demanding $t_{SL} = 250$ μs or more (dashed green line in Fig. 2d).

In an ideal SCC protocol configured with the optimum ionization pulse length, one expects the contrast to be dependent solely on the relative fraction of NVs left in the negatively charged state, and hence nearly insensitive to the readout pulse duration (at least in the limit case where the readout power has been chosen to produce negligible NV⁻ ionization, see Fig. 1b). As shown in Fig. 2c, however, this is not the case in our experiments, fundamentally because, as implemented, the SCC protocol preferentially ionizes NVs whose spin state

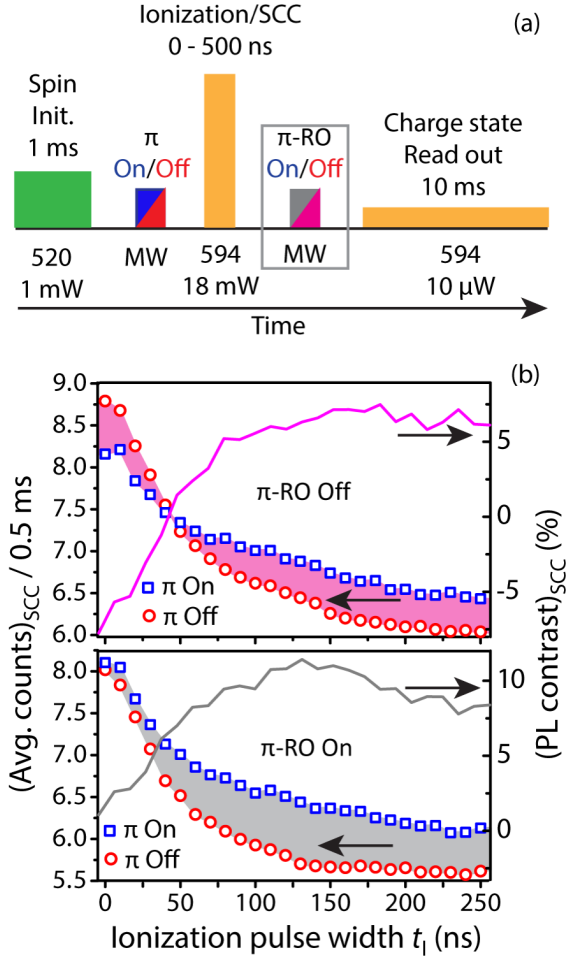

longer readout times.

The advantages of the SCC protocol can be best evaluated by calculating the speedup in the measurement time required to achieve a target sensitivity. In general, we can calculate the speedup factor from the formula

$$\text{Speedup Factor} = \left(\frac{\eta_{CSR}}{\eta_{SCC}}\right)^2,$$

where $\eta_{SCC}$ and $\eta_{CCR}$ are the sensitivities for spin-to-charge and conventional spin readout, respectively. For our 10 µW readout laser and a 20 ms readout time, a tenfold reduction (or better) in measurement time is possible when the measurement sequence length is 3 ms (or longer) (Fig 4a). In our present experiments, the speedup factor is limited by the readout laser power and pulse width. The speedup factor (and hence the detection sensitivity) can be increased by reducing the readout time, which, of course, demands increased readout laser powers. Given the quadratic growth of the NV⁻ ionization rate with laser power at 594 nm, we scale the readout time from the formula

$$\text{Readout Time} = T_0 \left(\frac{P_0}{P}\right)^2,$$

where $T_0 = 20$ ms and $P_0 = 10$ µW respectively denote the readout time and laser power used herein. The projected change in sensitivity is shown in Fig 4b for up to a tenfold increase in readout power (P). Compared to our present experimental

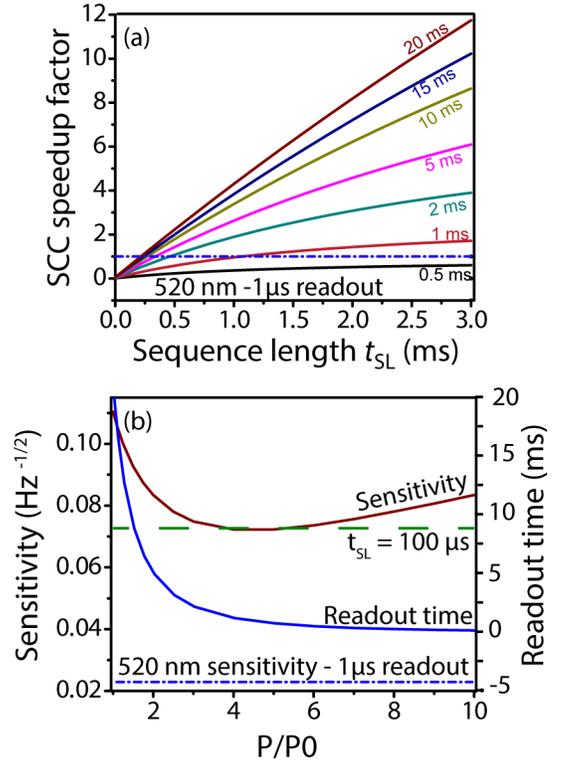

**Figure 3**: Improving SCC contrast with a readout π-pulse (π-RO). (a) Pulse sequence for comparing readouts with and without a π-RO (highlighted with a dotted square). (b) Gray (pink) shaded regions bound by the average counts of 0.5 ms SCC readout with (without) MW π-RO.

immediately prior to the ionization pulse is $m_S = 0$. At short readout times, therefore, the contrast is less than optimum because NVs initially in the $m_S = \pm 1$ states must be spin-polarized before the fluorescence reaches a maximum. This problem can be circumvented through the addition of a second MW pulse transferring the $m_S = \pm 1$ populations into the $m_S = 0$ state immediately before readout (Fig. 3a). A demonstration is presented in Fig. 3b for a 0.5 ms readout: The upper and lower plots allow us to compare the average spin-dependent count rates and resulting contrast for an SCC sequence with and without the additional readout MW pulse (denoted as π-RO in Fig 3a). Without the π-RO pulse and for ionization pulses of up to about 50 ns, the count rate is higher when the spin state prior to ionization is $m_S = 0$ (no preceding MW pulse is applied); the trend is inverted at longer times, i.e. the count rate from NV⁻ spins initialized into $m_S = \pm 1$ ultimately becomes dominant (upper half in Fig. 3b). This crossing between PL traces with different initial spin states effectively reduces the contrast and hence negatively impacts the resulting SNR. As shown in the lower half of Fig. 3b, the addition of a π-RO pulse immediately prior to the readout pulse removes this problem, allowing us to reach a contrast of 12%, on par with the contrast achieved with

**Figure 4**: (a) Measurement speedup factor for SCC, calculated for various readout times. The dash-dot line corresponds to the speedup factor of 1, which is the 'break even' point with standard readout. (b) Improvement in sensitivity calculated by increasing the readout power P relative to $P_0 = 10$ µW, used herein. The corresponding readout times are labelled on the right vertical axis. The dash-dot and dashed lines are sensitivities for 520 nm, 1µs readout assuming sequence lengths $t_{SL} = 0$ and 100 µs, respectively.

conditions, we find that the break-even point could be reduced to a sequence length of 100 μs for a 4× increase in readout power

In conclusion, we have demonstrated spin to charge conversion of NV ensembles in a bulk diamond sample. For the current experimental conditions, the SCC readout outperforms conventional readout for sequences longer than ~250 μs, limited by the readout and charge initialization laser powers. Additional sensitivity improvements could be attained through samples engineered to host a greater NV concentration, particularly those where the NV-to-nitrogen ratio is higher. Complementing prior studies, the observations reported herein should prove relevant to applications where NV spin ensembles in bulk crystals are important, particularly in sensing or imaging geometries where the target system sits on a diamond substrate engineered to host multiple NVs[11,25,26].

All authors acknowledge support from the National Science Foundation through grants NSF-1619896 and NSF-1547830, and from Research Corporation for Science Advancement through a FRED Award.

[1] G. Balasubramanian, I.Y. Chan, R. Kolesov, M. Al-Hmoud, J. Tisler, C. Shin, C. Kim, A. Wojcik, P.R. Hemmer, A. Krueger, T. Hanke, A. Leitenstorfer, R. Bratschitsch, F. Jelezko, and J. Wrachtrup, Nature **455**, 648 (2008).

[2] J.R. Maze, P.L. Stanwix, J.S. Hodges, S. Hong, J.M. Taylor, P. Cappellaro, L. Jiang, M.V.G. Dutt, E. Togan, A.S. Zibrov, A. Yacoby, R.L. Walsworth, and M.D. Lukin, Nature **455**, 644 (2008).

[3] J.-P. Tetienne, T. Hingant, L. Rondin, A. Cavaillès, L. Mayer, G. Dantelle, T. Gacoin, J. Wrachtrup, J.-F. Roch, and V. Jacques, Phys. Rev. B **87**, 235436 (2013).

[4] F. Dolde, H. Fedder, M.W. Doherty, T. Nöbauer, F. Rempp, G. Balasubramanian, T. Wolf, F. Reinhard, L.C.L. Hollenberg, F. Jelezko, and J. Wrachtrup, Nat. Phys. **7**, 459 (2011).

[5] M.W. Doherty, V.V Struzhkin, D.A. Simpson, L.P. McGuinness, Y. Meng, A. Stacey, T.J. Karle, R.J. Hemley, N.B. Manson, L.C.L. Hollenberg, and S. Prawer, Phys. Rev. Lett. **112**, 47601 (2014).

[6] V.M. Acosta, E. Bauch, M.P. Ledbetter, A. Waxman, L.-S. Bouchard, and D. Budker, Phys. Rev. Lett. **104**, 70801 (2010).

[7] N. Bar-Gill, L.M. Pham, A. Jarmola, D. Budker, and R.L. Walsworth, Nat. Commun. **4**, 1743 (2013).

[8] L Rondin and J-P Tetienne and T Hingant and J-F Roch and P Maletinsky and V Jacques, Reports Prog. Phys. **77**, 56503 (2014).

[9] H. Clevenson, M.E. Trusheim, C. Teale, T. Schröder, D. Braje, and D. Englund, Nat. Phys. **11**, 393 (2015).

[10] T. Wolf, P. Neumann, K. Nakamura, H. Sumiya, T. Ohshima, J. Isoya, and J. Wrachtrup, Phys. Rev. X **5**, 41001 (2015).

[11] D.A. Simpson, R.G. Ryan, L.T. Hall, E. Panchenko, S.C. Drew, S. Petrou, P.S. Donnelly, P. Mulvaney, and L.C.L. Hollenberg, Nat. Commun. **8**, 458 (2017).

[12] H.J. Mamin, M. Kim, M.H. Sherwood, C.T. Rettner, K. Ohno, D.D. Awschalom, and D. Rugar, Science **339**, 557 LP (2013).

[13] I. Lovchinsky, A.O. Sushkov, E. Urbach, N.P. de Leon, S. Choi, K. De Greve, R. Evans, R. Gertner, E. Bersin, C. Müller, L. McGuinness, F. Jelezko, R.L. Walsworth, H. Park, and M.D. Lukin, Science **351**, 836 LP (2016).

[14] L. Thiel, D. Rohner, M. Ganzhorn, P. Appel, E. Neu, B. Müller, R. Kleiner, D. Koelle, and P. Maletinsky, Nat. Nanotechnol. **11**, 677 (2016).

[15] J.-P. Tetienne, T. Hingant, J.-V. Kim, L.H. Diez, J.-P. Adam, K. Garcia, J.-F. Roch, S. Rohart, A. Thiaville, D. Ravelosona, and V. Jacques, Science **344**, 1366 LP (2014).

[16] M. Pelliccione, A. Jenkins, P. Ovartchaiyapong, C. Reetz, E. Emmanouilidou, N. Ni, and A.C. Bleszynski Jayich, Nat. Nanotechnol. **11**, 700 (2016).

[17] E. van O. and N.B.M. and M. Glasbeek, J. Phys. C Solid State Phys. **21**, 4385 (1988).

[18] B.J. Shields, Q.P. Unterreithmeier, N.P. De Leon, H. Park, and M.D. Lukin, Phys. Rev. Lett. **114**, 136402 (2015).

[19] D.A. Hopper, R.R. Grote, A.L. Exarhos, and L.C. Bassett, Phys. Rev. B **94**, 241201 (2016).

[20] H. Jayakumar, J. Henshaw, S. Dhomkar, D. Pagliero, A. Laraoui, N.B. Manson, R. Albu, M.W. Doherty, and C.A. Meriles, Nat. Commun. **7,** 12660 (2016).

[21] S. Dhomkar, P.R. Zangara, J. Henshaw, and C.A. Meriles, Phys. Rev. Lett. **120**, 117401 (2018).

[22] D.A. Hopper, R.R. Grote, S.M. Parks, and L.C. Bassett, ACS Nano **12**, 4678 (2018).

[23] S. Dhomkar, H. Jayakumar, P.R. Zangara, and C.A. Meriles, Nano Lett. **18,** 4046 (2018).

[24] N. Aslam, G. Waldherr, P. Neumann, F. Jelezko, and J. Wrachtrup, New J. Phys. **15**, 13064 (2013).

[25] J.F. Barry, M.J. Turner, J.M. Schloss, D.R. Glenn, Y. Song, M.D. Lukin, H. Park, and R.L. Walsworth, Proc. Natl. Acad. Sci. **113**, 14133 LP (2016).

[26] E. Schäfer-Nolte, L. Schlipf, M. Ternes, F. Reinhard, K. Kern, and J. Wrachtrup, Phys. Rev. Lett. **113**, 217204 (2014).